\RequirePackage{fix-cm}
\documentclass[smallextended]{svjour3}       
\smartqed  

\newcommand{\vect}[1]{{\mathbf{#1}}}
\newcommand{\norm}[1]{\| #1 \|}
\def \arg{\hbox{arg\ }}

\def\booknames#1/#2/#3/#4{#1:#2:#3:#4}
\usepackage{amsmath}
\usepackage[nothing]{algorithm}
\usepackage{algorithmic}
\usepackage{graphicx}

\usepackage{color}
\usepackage{soul}

\usepackage{svn}
\SVN $Author: diego $
\SVN $Date: 2012-12-03 14:38:51 -0500 (lun., 03 déc. 2012) $
\SVN $Rev: 897 $
\begin{document}

\title{ Dual-Based Bounds for Resource Allocation in
Zero-forcing Beamforming OFDMA-SDMA Systems}


\titlerunning{Dual-Based Bounds for Resource Allocation in
ZF OFDMA-SDMA Systems}        

\author{Diego Perea-Vega \and Andr\'e Girard \and Jean-Fran\c{c}ois
Frigon}

\institute{D. Perea-Vega$^*$, J.F. Frigon \at Department of
Electrical Engineering \'Ecole Polytechnique de Montr\'eal C.P.
6079, succ.
centre-ville, Montr\'eal, QC, Canada, H3C 3A7 \\
Tel: 1-514-340-4711 ext. 3642 Fax: 1-514-340-5892 \\
              \email{enrique.perea@polymtl.ca, j-f.frigon@polymtl.ca}           
\and
A. Girard \at GERAD Group for Research in Decision Analysis \\
3000, chemin de la C\^ote-Sainte-Catherine,
Montr\'eal, QC, Canada, H3T 2A7 \\
              \email{andre.girard@gerad.ca}             \\
$^*$Corresponding author
 }

\date{Received: date / Accepted: date}

\maketitle

\begin{abstract}
We consider multi-antenna base stations using orthogonal
fre\-quen\-cy-di\-vision multiple access and space division
multiple access techniques to serve single-antenna users.
Some users, called  real-time users, have minimum rate
requirements and must be served in the current time slot while
others, called  non real-time users, do not have strict timing
constraints and are served on a best-effort basis. The resource
allocation problem is to find the assignment  of users to
subcarriers and the transmit beamforming vectors that maximize
the total user rates subject to power and
minimum rate constraints. In general, this is a nonlinear and non-convex
program and the zero-forcing technique used here makes it integer as
well, exact optimal solutions cannot
be computed in reasonable time for realistic cases. For this reason,
we present a technique to compute both upper and lower bounds and show
that these are quite close for some realistic cases.

First, we formulate the dual problem whose optimum provides an
upper bound to all feasible solutions. We then use a simple method
to get a primal-feasible point starting from the dual optimal
solution, which is a lower bound on the primal optimal solution.
Numerical results for several cases show that the two bounds are
close so that the dual method can be used to benchmark any
heuristic used to solve this problem. As an example, we provide
numerical results showing the performance gap of the well-known
weight adjustment method and show that there is considerable room
for improvement.
\end{abstract}
%
\section {Introduction} \label{sec:Introduction}
With the ubiquitous use of smart phones, tablets, laptops and
other devices, traffic demand on wireless access networks is
increasing geometrically. Multi-antenna base stations using
orthogonal frequency-division multiple access (OFDMA) and space
division multiple access (SDMA) can transmit at the same time to
different sets of users on multiple subcarriers. This diversity increases the system
throughput by assigning transmitting resources to users with good
channel conditions. High data rates are thus made possible by
exploiting the degrees of freedom of the system in time, frequency
and space dimensions. OFDMA-SDMA is also supported by WiMAX and
LTE-Advanced systems, which are the technologies currently being set
up  to
implement the fourth generation (4G) cellular
networks~\cite{lte09a,wim09a}.

Due to these increased degrees of freedom, it is critical to use a
dynamic and efficient scheduling and resource allocation (RA) mechanism that
takes full advantage of all OFDMA-SDMA transmitting
resources~\cite{letaief06}.
In this mechanism, the scheduler selects the users that are
served at each frame and the RA algorithm  allocates to these
users the transmission resources required to meet the quality of
service (QoS) requested from the upper layers.

In this paper, we focus on the RA part of the problem for an
OFDMA-SDMA system supporting real time traffic with minimum rate
requirements. We propose a dual-based method to get an upper bound
to all feasible solutions. Because the dual solution is not
necessarily primal feasible, we also present an algorithm to get a
feasible point from the dual solution, which is a lower bound to
the optimal primal solution. The importance of these bounds is
that they give us limits on the duality gap and we can use them to
estimate how far the solution given by any heuristic method is
from the optimum.
%
\subsection{ State of the Art}
\label{sec:state_art}
The general resource allocation is a nonlinear, non-convex program so that
it is almost impossible to solve it directly
for any realistic number of subcarriers, users and antennas.
For this reason, most research work focuses on
developing heuristic algorithms. In this context, an important
question  is always how accurate are the results. For the resource
allocation problem with rate constraints, it turns out that there are very few results of
that kind, as we shall see.

Traffic in the system can be divided into two main groups:
delay-sensitive real time (RT) services and delay-insensitive non
real time (nRT) services. Early work on OFDMA systems focused on
solving
the RA problem for nRT services only, where the objective is to
maximize the total throughput with only power constraints and
possibly minimum BER constraints. In~\cite{bar07}, the complete
optimization problem is divided into two sub-problems: Selection of
users for each carrier and then power allocation to these users, which are both
solved by a heuristic. A similar approach using zero-forcing (ZF)
beamforming is reported in \cite{maciel07}. The  work
of~\cite{bar07,maciel07} does not solve the complete optimization
problem; instead, it separates it into uncoupled subproblems that
provide sub-optimal solutions.

There is a definite need to benchmark the
performance of these heuristic algorithms.
For the RA problem with
nRT traffic only, several methods have been proposed to compute
near-optimal solutions.
For example, genetic algorithms are proposed in
\cite{ozbek09}  while \cite{tsang04,chan07,xingmin10,perea10}
provide methods to compute a near-optimal solution based on dual
decomposition. In addition to providing a benchmark, near-optimal
algorithms can also lead to the design of efficient RA methods as
shown in~\cite{perea10}, where heuristic algorithms derived from
the dual decomposition method are proposed.

Several methods have been used to solve the RA problem for
OFDMA-SDMA systems with both RT and nRT traffic.  In
{\cite{tsai08}}, the objective is to maximize the sum of the user
rates subject to per-user minimum rate constraints that model the
priority assigned to each user at each frame. The optimization
problem is solved approximately for each frame by minimizing a
cost function representing the increase in power needed when
increasing the number of users or the modulation order.
The advantages of this approach are that it handles user
scheduling and resource allocation together  and supports
Rt and nRT traffic.
Its weaknesses are that no comparison
is made against a near-optimal solution and the method used to
determine user priorities at every frame is very complex.

In \cite{chung09}, both RT and nRT traffic are supported.
Priorities are set according to the remaining deadline time for RT
users and to the difference between the achieved rate and the
desired rate required for nRT users. The user with the highest
priority is paired with the subchannel with the highest vector
norm and semi-orthogonal users are multiplexed  on the same
subchannel. Comparisons against the algorithm of \cite{tsai08}
show that the packet drop rate for RT users is significantly
reduced.  The algorithm's complexity is also reduced because of
the semi-orthogonality criteria used to add users. However, as in
{\cite{tsai08}}, a performance comparison with a near-optimal
solution is not provided.

In \cite{huang08}, the objective is to
maximize a utility function without any hard minimum rate
constraints for the RT users. The channel quality information is
added to the utility function to favor users with good channel
conditions and priorities are set by increasing user weights in
the utility function. The advantage is that the
per-frame optimization problem has only a power constraint and no
rate constraints, which makes it simpler to solve. The
disadvantage with this \emph{reactive} approach is that RT users
with poor channel conditions are backlogged until their delay is
close to the deadline, increasing the average delay and jitter.

In \cite{papoutsis10}, a heuristic algorithm is proposed for the
sum rate maximization problem with proportional rates among the
user data rates, i.e.,  the ratio among allocated user rates is
predetermined. The criteria to form user groups includes
semi-orthogonality as in \cite{chung09}, but also fairness through
proportional rate constraints. This method is extended to include
hard minimum rates in \cite{vasileios11} which attempts to solve
exactly the same problem we formulate in section
\ref{sec:SysDescription}. Again, there is not reported method to
evaluate the accuracy of these heuristics, except by comparing
them to each other.

Another approach is to maximize the sum rate subject to
constraints on the average rate delivered to a
user~\cite{tralli11}. However, unlike the work presented
in~\cite{wang-giannakis08} where an optimal solution is provided
for the single antenna OFDMA RA problem with average rate
constraints, the algorithm presented in~\cite{tralli11} is a
heuristic approximation. Note also that with average rate
constraints, RT users tend to be served when they have good
channel conditions which can create unwanted delay violations and
jitter.
\subsection{Paper Contribution and Organization}
None of the previous work provide a near-optimal solution to the
OFDMA-SDMA RA problem with minimum rate requirements. This is
important not only as a benchmark for any heuristic
algorithms, but also to get a better insight into the problem and
to help devise efficient heuristics.

The main contribution of this paper is  a method that provides an
upper bound to the following OFDMA-SDMA RA problem for mixed RT
and nRT traffic:  For a given time slot, find the user selection
and beamforming vectors that maximize a linear function of
the users rates, given a total transmit power constraint and
minimum rate constraints for RT users. The user weights in the
linear utility function are arbitrary and can be the result of a
prioritization or fairness policy by the scheduler. We solve the
RA problem by  Lagrange decomposition and we show, for small cases
where we can find the optimal solution, that the duality gap is
small.

A second contribution is  a simple off-line heuristic algorithm to
compute a feasible point based on the dual solution. This point is
a lower bound for the optimal solution. This lower bound, in
conjunction with the upper bound from the dual, can be used to
limit the optimality gap in larger cases where an optimal solution
is not available.

We then
study several cases where we compare the performance of the
upper and lower bounds. The results show that the two
bounds are tight when the number of RT users is small and that
their difference increases for larger values but that
it stays quite small. Thus, the dual method
provides a good approximation to the optimal solution. We also
compare the performance of the weight adjustment algorithm
versus the solution provided by the two bounds. The results
indicate that adjusting the user weights to prioritize RT users
can lead to significantly sub-optimal solutions.

We describe the system and formulate the optimization problem we
want to solve in Section~\ref{sec:SysDescription} where we also
briefly discuss a direct enumeration method to find the optimal
solution for small problems. We present in
Section~\ref{sec:dual-based-solution} the dual method and
two algorithms: One that finds the dual solution (the upper bound)
and the other that finds a dual-based primal feasible solution
(the lower bound).
In Section~\ref{sec:results}, we present numerical results showing
the accuracy of the upper and lower bounds and of the weight
adjustment algorithm for different scenarios. Finally, we present
our conclusions in Section~\ref{sec:conclusions} .
\section {System Description and Problem Formulation}
\label{sec:SysDescription}
We consider the resource allocation problem for the downlink
transmission in a MISO system with a single base station. There are $K$
users, some of which have RT traffic with minimum rate
requirements while the others have nRT traffic that can be served on a best-effort basis. The BS
is equipped with $M$ transmit antennas and each user has one
receive antenna. In this configuration, the BS can transmit in the downlink data
to different users on each subcarrier by performing
linear transmit beamforming precoding. At each OFDM symbol, the
BS can change the beamforming vector for each user on
each subcarrier to maximize some performance function. In this
paper, we assume that we use a channel coding that reaches the channel
capacity. The data rate are in units of bits per OFDM
symbol, or equivalently bits per second per Hertz (bps/Hz).

\subsection{Signal Model} \label{subsec:SignalModel}
First, we describe the model used to compute the bit rate received
by each user. Define
\begin{description}
\item[$\tilde{s}_{k,n}$] the symbol transmitted by the BS to user $k$ on
subcarrier
  $n$. We assume that the $\tilde{s}_{k,n}$ are independently identically
  distributed (i.i.d) random variables
with $\tilde{s}_{k,n} \sim \mathcal{CN}(0,1)$.
\item[$\vect{w}_{k,n}$] an $M$-component column vector
representing the beamforming vectors for user $k$ on subcarrier $n$.
Unless otherwise noted, we denote $\vect{w}$ the vector made up by
the column stacking of the vectors
 $\vect{w}_{k,n}$.
\item[$\mathbf{x}_{n}$] an $M$-component column vector
representing the signal sent by the array of $M$ antennas at the
BS  for each subcarrier $n$.
\item[$\mathbf{h}_{k,n}$] an $M$-component row vector representing
the channel between the $M$ antennas at the BS and the receive
antenna at user $k$  for each subcarrier $n$.
\item[$z_{k,n}$] the additive white gaussian noise at the receiver
for  user $k$ on subcarrier $n$. The $z_{k,n}$ are i.i.d. Gaussian
random variables and, without loss of generality, we assume that
they have unit variance, that is $z_{k,n}  \sim
\mathcal{CN}(0,1)$.
\item[$y_{k,n}$] the signal received by user $k$ on subcarrier
$n$.
\item[$r_{k,n}^0$] the rate of user $k$ on subcarrier $n$ in bps/Hz.
\end{description}
The signal vector $\vect{x}_{n}$ is built by a linear precoding
scheme which is a linear transformation of the information symbols
$\tilde{s}_{k,n}$:
\begin{equation}
\label{eq:x_signal} \vect{x}_{n}= \sum_{k=1}^K \vect{w}_{k,n}
\tilde{s}_{k,n}.
\end{equation}
The signal received by user $k$ on subcarrier $n$ is then given by
\begin{align}
y_{k,n}&= \vect{h}_{k,n} \vect{x}_{n}+z_{k,n} \notag\\
&=\vect{h}_{k,n} \vect{w}_{k,n} \tilde{s}_{k,n} + \sum_{j \neq k}
\vect{h}_{k,n} \vect{w}_{j,n} \tilde{s}_{j,n} + z_{k,n}.
\label{eq:y_k_rx}
\end{align}
The second and third terms in the right-hand side of (\ref{eq:y_k_rx})
correspond to the interference and noise terms.
Since the signals and noise are Gaussian, their sum
is also Gaussian and the data rate of user $k$ for
subcarrier $n$ is given by the Shannon channel capacity for an
additive white Gaussian noise channel:
\begin{equation}
\label{eq:bitrate} r_{k,n}^0\left(\vect{w}\right)= \log_2 \left( 1
+ \frac {{|\vect{h}_{k,n} \vect{w}_{k,n}|}^2} {1 + \sum_{j \neq k}
  {|\vect{h}_{k,n} \vect{w}_{j,n} |}^2 } \right).
\end{equation}
%
\subsection{Rate Maximization Problem}
\label{sec:rate-maxim-probl}

The general rate maximization problem corresponding to the
OFDMA-SDMA RA problem with mixed RT and nRT traffic is to find a
set of beamforming vectors $\vect{w}_{k,n}$ that will maximize the
weighted sum of user rates. This is limited by the total power
available for the transmission at the base station and some users
with real time QoS requirements must receive a minimum rate. More
precisely, we assume that we know
\begin{description}
\item[$K$] Number of users in the cell. \item[$\mathcal{K}$] Set
of users in the cell: $\{1, \ldots, K\}$.
\item[$\mathcal{D}$]  A subset of $\mathcal{K}$ containing the
users that have minimum rate requirements. We define  $D= \vert
\mathcal{D} \vert $.
\item[$\check{d}_{k}$] Minimum rate requirement for user
  $k$.
\item[$M$] Number of antennas at the BS.
 \item[$N$] Number of subcarriers available.
\item[$\check{P}$] Total power available at the base station for
transmitting over all channels.
\item[$c_k$] Weight of the user rates in the objective function.
These could be computed by the scheduler to
implement prioritization or fairness.
\end{description}
We then want to solve the following optimization problem to obtain
the resource allocation
%
\begin{align}
\label{eq:gral_form} \max_{ \vect{w} }  \sum_{n=1}^{N}
\sum_{k=1}^{K}&
c_k r_{k,n}^0 ( \vect{w} )\\
\sum_{n=1}^{N} \sum_{k=1}^{K} \norm{ {\vect{w}}_{k,n} }^2 & \leq
\check{P} \label{eq:powergen} \\
\sum_{n=1}^{N} r_{k,n}^0 ( \vect{w} ) & \ge \check{d}_{k} , \quad
\forall k \in { \mathcal{D} }. \label{eq:minrategen}
\end{align}
The total transmit power is represented by the sum of the $L_2$
norms of the beamforming vectors in
constraint~(\ref{eq:powergen}). The achievable rate over all
subcarriers should be higher or equal than the required minimum
rate per user as in~(\ref{eq:minrategen}).

Problem~(\ref{eq:gral_form}--\ref{eq:minrategen}) is a non-convex,
non-linear optimization problem. Using an exact algorithm to find
a global optimal solution is very hard considering the size of a
typical problem where there can be up to a hundred users and
hundreds of sub-channels. Another option is to use a standard
non-linear program (NLP) solver to compute a local optimal
solution and use different starting points in the hope of finding
a good global solution. The problem with this approach is that 1)
we don't know how close we are to the true optimum and 2) the
technique is quite time-consuming.

Nevertheless, we tried this
approach for some small cases and observed that most users end up
with a zero beamforming vector and only a small subset of users
actually get some rate, often  no more than $M$. Furthermore, in
accordance to what was reported for the SDMA problem
in~\cite{yoo06}, we observed that at high SNR,  a so-called
\emph{zero-forcing} (ZF) solution is very close to the local
optimum. This ZF solution can be easily computed by channel
diagonalization and water-filling power allocation and is
near-optimal compared to the general beamforming solution in the
high SNR regime. Moreover, due to multi-user diversity, there is a
good chance of finding  users with high SNR channels when the total number
of users increase. Therefore, in a multi-user Rayleigh fading
environment, the ZF beamforming technique can provide results
close to the general beamforming solution, even in the moderate SNR
regime. For these reasons, we now turn to the so-called
Zero-Forcing beamforming strategy.
%
\subsection{Zero-Forcing Beamforming}
\label{subsec:ZF_stra}
In general, user $k$ is subject to the interference from other
users which reduces its bit rate, as indicated by the denominator
in~(\ref{eq:bitrate}).  Zero-forcing beamforming is a
strategy that completely eliminates interference from other users.
For each subcarrier $n$, we choose a set $s$ of $g \le M$ users
which are allowed to transmit. This is called an \emph{SDMA} set.
We then impose the condition that for each user $k$ in this set,
the beamforming vector of user $k$ must be orthogonal to the
channel vectors of all the other users of the set. This amounts to
adding the orthogonality constraints
\begin{equation}
\label{eq:ortho_const} \vect{h}_{k,n} \vect{w}_{j,n} = 0 \quad j
\not= k,\, \, j, k \in s
\end{equation}
and the user $k$ data rate for subcarrier $n$ then simplifies to:
\begin{equation}
\label{eq:zfbitrate} r_{k,n}^0\left(\vect{w}_{k,n}\right)= \log_2
\left( 1 + {|\vect{h}_{k,n} \vect{w}_{k,n}|}^2 \right).
\end{equation}
With ZF beamforming, the problem is now made up of two parts. We
need to select a SDMA set $s(n)$ for each subcarrier $n$ and, for
each selected SDMA set, we must compute the beamforming vectors in
such a way that the total rate received by all users is maximized.
Because of this, we need to add another set of decision variables
\begin{description}
\item[$\alpha_{k,n}$]  a binary variable that is 1 if we allow
user
  $k$ to transmit on subcarrier $n$ and zero otherwise. We denote the collection of
  $ \alpha_{k,n} $ by
  the vector $\boldsymbol{\alpha}$ .
\end{description}
This results in the following ZF problem
%
\begin{align}
\max_{ \vect{w}, \boldsymbol{\alpha} } \sum_{n=1}^{N} \sum_{k=1}^{K} &
c_k r_{k,n}^0 ( \vect{w}_{k,n} )  \label{eq:objzf}\\
\sum_{n=1}^{N} \sum_{k=1}^{K} \norm{ \vect{w}_{k,n} }^2 & \leq
\check{P} \label{eq:powerz} \\
\sum_{n=1}^{N} r_{k,n}^0 ( \vect{w}_{k,n} ) & \ge \check{d}_{k} ,
\quad \forall k \in { \mathcal{D} } \label{eq:minratezf} \\
\sum_{k=1}^{K} \alpha_{k,n} & \le M,  \quad \forall n \label{eq:sdmasize} \\
| \vect{h}_{k,n} \vect{w}_{j,n} |^2 & \le B \left[ ( 1 -
\alpha_{k,n}) +  ( 1 - \alpha_{j,n} ) \right],  \quad \forall n,\,
\forall k,\, \forall j, \, k \not= j \label{eq:zfconst} \\
\norm{\vect{w}_{k,n}} & \le A \alpha_{k,n} \label{eq:zfub}\\
\alpha_{k,n} & \in \left\{ 0, 1 \right\} \label{eq:alphabin}
\end{align}
%
where $A$ and $B$ are
some large positive constants. Constraint~(\ref{eq:sdmasize})
guarantees that we do not choose more than $M$ users for each
subcarrier, constraint~(\ref{eq:zfconst}) guarantees that if we
have chosen two users $k$ and $j$, they meet the ZF constraints
and is redundant for other users, and
constraint~(\ref{eq:zfub}) guarantees that the beamforming vector
is zero for users that are not chosen.
Problem~(\ref{eq:objzf}--\ref{eq:zfub}) is a non-linear mixed
integer program (MIP) and these are known to be very hard to solve
exactly.

In this paper, whenever we need to get an exact solution,
we use a complete enumeration of the binary variables
$\boldsymbol{\alpha}$ satisfying the
constraints~(\ref{eq:sdmasize}). For each $\boldsymbol{\alpha}$,
we then  find the beamforming vectors maximizing the utility
function given the power, minimum rate and ZF constraints. The
optimal solution is the vector $\boldsymbol{\alpha}$ that
maximizes the utility function. The problem of finding the optimal
$\vect{w}$ variables for a particular $\boldsymbol{\alpha}$ is a
relatively small non-convex problem. Using the pseudo-inverse
approach to satisfy the ZF constraint (see
Section~\ref{sec:appr-solut-beamf}), it can also be transformed
into an approximate convex problem  and solved using a dual algorithm
similar to the one described in Section~\ref{sec:comp-dual-funct}.

It would seem that the zero-forcing model is not improving things
much: We have gone from a non-convex nonlinear program to a
non-convex mixed nonlinear program that can be solved by picking
the global solution from a large collection of small convex
problems. However, as we explain in section~\ref{sec:dual-based-solution}, this allows
us to design an efficient and accurate algorithm.
\section{Dual-Based Solution Method}
\label{sec:dual-based-solution}
Obviously, we cannot solve problem~(\ref{eq:objzf}--\ref{eq:alphabin}) fast
enough to use it in  real time. Not only
is it NP-complete~\cite{bar07} but the actual computation time becomes quickly
prohibitive for realistic sizes, even for off-line computations.
Still, we need
to compute solutions so that we can use them as benchmarks to
evaluate the quality of real time heuristic approximations. We now
present two off-line  solution techniques that are tractable for
problems of realistic size  based on the Lagrange
relaxation of the primal. Because
problem~(\ref{eq:objzf}--\ref{eq:alphabin}) is \emph{not} convex,
there will often be a strictly positive duality gap at the solution of the
dual. However, if it is small enough, we can use the solution provided by
the dual method as a useful benchmark to check the accuracy of
heuristic methods. Results in section \ref{sec:results} show that
in many cases, the duality gap is in fact less than a few percent.

Solving the ZF problem will require some form of search over the
$\alpha_{k,n}$ variables. Note that this ranges over all subsets
with a number of users smaller than or equal to $M$, so that the search
space is going to be fairly large. Our first transformation is
thus to separate the problem into single-subcarrier subproblems.
For this, we dualize the constraints~(\ref{eq:powerz})
and~(\ref{eq:minratezf}) since they are the ones that couple the
subcarriers. Define the dual variables
\begin{description}
\item[$\lambda$]  Lagrange multiplier for power
constraint~(\ref{eq:powerz}). \item[$\mu_k $]  Lagrange
multipliers for minimum rate
  constraint~(\ref{eq:minratezf}) of   user $k$. The collection of
  $\mu_k$ is denoted $\boldsymbol{\mu}$.
\end{description}
In order to simplify the derivation, we also define the dual variables
$\mu_k$ for all users $k \in \mathcal{K}$. For users with no
minimum rate requirements ($k \notin \mathcal{D}$), we have $\mu_k
= 0 $ and $\check{d}_{k}=0$. In what follows, we use the standard form of Lagrangian
duality which is expressed in terms of minimization  with
inequality constraints of the form $\le$. Under these conditions,
the multipliers $\lambda, \boldsymbol{\mu} \ge 0$. We get the
partial Lagrangian
\begin{align}
  \mathcal{L} = - & \sum_{n=1}^{N} \sum_{k=1}^{K} c_k r_{k,n}^0 ( \vect{w}_{k,n}  )
  + \lambda \left[  \sum_{n=1}^{N} \sum_{k=1}^{K} \norm{ \vect{w}_{k,n} }^2  - \check{P}  \right]
  \nonumber \\  + & \sum_{k=1}^K  \mu_k \left[ \sum_{n=1}^{N}
    - r_{k,n}^0 ( \vect{w}_{k,n} )  + \check{d}_{k} \right]
  \nonumber \\
  = - &\lambda  \check{P} + \sum_{k=1}^K   \mu_k \check{d}_{k} +
  \sum_{n=1}^N \left\{  - \sum_{k=1}^K  (c_k+\mu_k) r_{k,n}^0 ( \vect{w}_{k,n}  )
    + \lambda \sum_{k=1}^K \norm{ \vect{w}_{k,n} }^2
\right\}
  \label{eq:partiallagr}
\end{align}
with constraints~(\ref{eq:sdmasize}--\ref{eq:alphabin}).
The value of the dual function $\Theta$ at some point $(\lambda,
\boldsymbol{\mu})$ is obtained by minimizing the Lagrange function
over the primal variables
\begin{equation}
  \Theta(\lambda, \boldsymbol{\mu} ) = \min_{\vect{w},
    \boldsymbol{\alpha} } \mathcal{L} ( \lambda, \boldsymbol{\mu} ,
  \vect{w}, \boldsymbol{\alpha} )
\label{eq:defdualf}
\end{equation}
and the dual problem is
\begin{align}
  \label{eq:defdualobj}
  \max_{\lambda, \boldsymbol{\mu} } \ & \Theta(\lambda, \boldsymbol{\mu} ) \\
  \lambda, \boldsymbol{\mu} & \ge 0 \label{eq:dualconst}
\end{align}
which we can solve by the well known subgradient
algorithm~\cite{bertsekas03}. From now on, we concentrate on the
calculation of the subproblem~(\ref{eq:defdualf}).
\subsection{Subchannel Subproblem}
\label{sec:subchsubpr}
Because of the relaxation of the carriers coupling constraints~(\ref{eq:powerz}--\ref{eq:minratezf}),
the subproblems in~(\ref{eq:defdualf}) decouple by subcarrier
since the objective~(\ref{eq:partiallagr}) is separable in $n$ and
so are constraints~(\ref{eq:sdmasize}--\ref{eq:zfub}). Computing
the dual function then requires the solution of $N$ independent
subproblems. For each subcarrier $n$,
this has the form
\begin{align}
  \label{eq:objsublagr}
  \min_{\vect{w}_n, \boldsymbol{\alpha}_n }  \,\,  - \sum_{k=1}^K  &( c_k + \mu_k
  )  r_{k,n}^0 ( \vect{w}_{k,n}  )
    + \lambda \sum_{k=1}^K \norm{ \vect{w}_{k,n} }^2 \\
\sum_k \alpha_{k,n} & \le M,  \label{eq:sdmasizelagr} \\
{\left| \vect{h}_{k,n} \vect{w}_{j,n} \right|}^2 & \le B \left[ ( 1 -
\alpha_{k,n}) +  ( 1 - \alpha_{j,n} ) \right], \quad \forall k, \,
\forall j,
\, k \not= j \label{eq:zfconstlagr}\\
\norm{\vect{w}_{k,n}} & \le A \alpha_{k,n} \label{eq:zfublagr} \\
\alpha_{k,n} & \in \left\{ 0, 1 \right\} \nonumber
\end{align}
where $\vect{w}_n$ is the vector made up by the column stacking of
the vectors $\vect{w}_{k,n}$ for subcarriers $n$ and
$\boldsymbol{\alpha}_n$ denote the collection of $\alpha_{k,n}$
for subcarrier $n$.
Problem~(\ref{eq:objsublagr}--\ref{eq:zfublagr}) is still a mixed
NLP, albeit of a smaller size.
\subsection{SDMA Subproblem}
\label{sec:sdma-subproblem}
A simple solution procedure is to enumerate all possible choices
for $\alpha_{k,n}$ that meet constraint~(\ref{eq:sdmasize}). This
is called the \emph{extensive} formulation of the problem. Each
choice defines a SDMA set $s$  and $\kappa = \vert s \vert$  and here,
we present a solution technique for the subproblems with given $s$.
For each $s$, the problem separates into
$\kappa$ independent problems to compute the optimal beamforming
vector $\vect{w}_{k,n,s}$ for each user  $k \in s$. For each user
$k \in s$ we know the set of channel vectors for the other members
of $s$ and we collect these vectors in the $(\kappa - 1) \times M$
matrix $\vect{H}_{k,n,s}$.
Problem~(\ref{eq:objsublagr}--\ref{eq:zfublagr}) can then be
rewritten as
%
%
\begin{align}
\min_s&\quad f_{n,s} \label{eq:subproblem} \\
f_{n,s}&=\sum_{k \in s} f_{k,n,s}^* \label{eq:theo1_2prove-b}
\end{align}
where $f_{k,n,s}^*$ is given by the solution of the optimization
\begin{align}
f_{k,n,s}^* & = \min_{\vect{w}_{k,n,s}} -c'_k \log_2 \left( 1 + |
\vect{h}_k
      \vect{w}_{k,n,s} |^2  \right) + \lambda \norm{\vect{w}_{k,n,s}}^2
\\
%
\vect{H}_{k,n,s} \vect{w}_{k,n,s} & = 0
\label{eq:orthosubpr}
\end{align}
%
where $c'_k = c_k + \mu_k$, $\vect{w}_{k,n,s}$ is the beamforming
vector for user $k$ on subcarrier $n$ for SDMA set $s$, and
$\vect{w}_{n,s}$ is the vector made up by the column stacking of
the vectors $\vect{w}_{k,n,s}$ for the $\kappa$ users in $s$.
Note that constraint~(\ref{eq:sdmasizelagr})  is automatically
satisfied by the construction of $s$,
constraint~(\ref{eq:zfublagr}) simply drops out since
$\vect{w}_{k,n,s} = 0$ for $ k \not\in s$ and
constraint~(\ref{eq:zfconstlagr}) remains only for $k \in s$, but
we write it as~(\ref{eq:orthosubpr}) because we are considering
only users that belong to SDMA set $s$.

This is certainly not a feasible real-time algorithm, but for
realistic values of $K$ and $M$ the number of
SDMA sets is still manageable and the optimization
sub-problem~(\ref{eq:subproblem}--\ref{eq:orthosubpr}) is a small
nonlinear program with $M$ variables and $\kappa - 1$ linear constraints. It
can thus be solved quickly by a number of techniques. Still, the
overall  computation load can be quite large. There will be
$\kappa$ such problems to solve for each SDMA set, and there are $
S =  \sum_{i=1}^M \binom{K}{i}$ such sets for each of the $N$
subcarriers so that we have to solve the problem $\kappa \times S
\times N$ times, and this for each iteration of the subgradient
algorithm. Clearly, any simplification of the beamforming
subproblem can reduce the overall computation time significantly.

%

%

%
\subsection{Approximate Solution to the Beamforming Problem}
\label{sec:appr-solut-beamf}
%
This can be done by the following construction. Instead of
searching in the whole orthogonal subspace  of $\vect{H}_{k,n,s}$ as
defined by~(\ref{eq:orthosubpr}), we pick a direction vector in
that subspace and search only on its support. This will give a
good approximation to the extent that the direction vector is
close to the optimal vector. The choice of direction is motivated
by the fact that the objective function depends only on the
product $\vect{h}_k \vect{w}_{k,n,s}$. We then introduce a new independent
variable
\begin{equation}
  \label{eq:defqsubpr}
  q_{k,n,s} = \vect{h}_k \vect{w}_{k,n,s}
\end{equation}
and because this variable is not independent of $\vect{w}_{k,n,s}$, we add
(\ref{eq:defqsubpr}) as a constraint. We then get the
equivalent problem
\begin{align}
  \label{eq:usersubpobeq}
  \max_{\vect{w}_{k,n,s}, q_{k,n,s} } & c'_k \log_2 \left( 1 + |q_{k,n,s}|^2  \right) - \lambda \norm{\vect{w}_{k,n,s}}^2 \\
  \vect{h}_k \vect{w}_{k,n,s}    & = q_{k,n,s} \label{eq:qwconstsub} \\
\vect{H}_{k,n,s} \vect{w}_{k,n,s} & = 0. \label{eq:usersubpceq}
\end{align}
Constraints~(\ref{eq:qwconstsub}) and (\ref{eq:usersubpceq}) can
then be rewritten in the standard form $\vect{G}_{k,n,s}
\vect{w}_{k,n}$ $=\vect{b}_{k,n,s}$ where the $\vect{G}_{k,n,s}$
matrix is the concatenation of $\vect{h}_k$ and $\vect{H}_{k,n,s}$
and $\vect{b}_{k,n,s} = [q_{k,n,s}, 0 ,0 \ldots 0]^T$.

Since we are proposing to transform the constrained optimization
over the $\kappa$ variables into an unconstrained optimization
over $q_{k,n,s}$ only, we must be able to express
$\vect{w}_{k,n,s}$ as a function of $q_{k,n,s}$. The linear system
being under-determined, this is obviously not unique. We propose
to use $\vect{G}^+_{k,n,s}$, the pseudo-inverse of
$\vect{G}_{k,n,s}$, for the back-transformation $\vect{w}_{k,n,s}
= \vect{G}^+_{k,n,s} \vect{b}_{k,n,s}$.
The pseudo-inverse picks the vector of
minimum norm compatible with the linear system. In other words,
choosing this transformation will \emph{minimize}
$\norm{\vect{w}_{k,n,s}}$ so that it is minimizing the power term
in the objective function in~(\ref{eq:usersubpobeq}). Because
$\lambda \ge 0$, this has the effect of contributing to the
maximization of $f_{k,n,s}^*$. Note that this technique provides
only an approximate solution of the beamforming problem; we cannot
invoke Theorem~1 from~\cite{wiesel07} which shows that in certain
cases, the pseudo-inverse transformation is optimal. A strong
assumption for the theorem is that the objective function depends
only on the $q_{k,n,s}$ variable, which is not the case here
since~(\ref{eq:usersubpobeq}) also depends on
$\norm{\vect{w}_{k,n,s}}^2$. However, we observed from numerical results that
the difference between the pseudo-inverse solution and the optimal
solution is small.
%
%
With this approximation we fix the direction of the beamforming
vectors to
\begin{align}
\vect{w}_{k,n,s} = & \vect{G}^+_{k,n,s}
\vect{b}_{k,n,s} \nonumber \\
 = & q_{k,n,s} [\vect{G}^+_{k,n,s}]_1   \nonumber
\end{align}
where $[\vect{G}_{k,n,s}^+]_1$ denotes the first column of
$\vect{G}_{k,n,s}^+$. Now, we can obtain a problem formulation in
terms of the user powers only by replacing the following
expression in (\ref{eq:usersubpobeq}):
\begin{equation}
  \norm{\vect{w}_{k,n,s}}^2 = \gamma_{k,n,s}^2 p_{k,n,s},
\end{equation}
where $ \gamma_{k,n,s} = \norm{ [\vect{G}_{k,n,s}^+]_1} $ and
$p_{k,n,s} = |q_{k,n,s}|^2$. Adding the constraint $p_{k,n,s} \ge
0$, we get the equivalent problem
\begin{align}
  \max_{p_{k,n,s}}  \ & c_k' \log (1 + p_{k,n,s}) - \lambda \gamma_{k,n,s}^2 p_{k,n,s}  \label{eq:notespinvobj}\\
  p_{k,n,s} & \ge 0
\end{align}
which has the solution
\begin{equation}
  \label{eq:solpseudo}
  p_{k,n,s} = \max \left\{  0,  \frac{c_k'}{\lambda \gamma_{k,n,s}^2} - 1 \right\}
\end{equation}
so that the computation time is basically that for computing
$\vect{G}_{k,n,s}^+$. Also note that using $\vect{G}_{k,n,s}^+$ we can also
find the optimal beamforming vectors for all users in $s$, the only
difference being that $ \gamma_{k,n,s}$ is computed using the column
of $\vect{G}_{k,n,s}^+$ corresponding to the channel vector of this
user.
\subsection{Solving the Dual Problem}
\label{sec:comp-dual-funct}
To summarize, the dual function $\Theta(\lambda, \boldsymbol{\mu})$ is obtained for the current values of the
multipliers by finding for each subcarrier $n=1,\dots,N$ the optimal SDMA set $s^*(n)$ to the minimization problem in~(\ref{eq:subproblem}), where
\begin{equation}
\label{eq:fns-def}
 f_{n,s} (\lambda, \boldsymbol{\mu})=-\sum_{k
\in s}\left[c'_k \log \left( 1 + p_{n,s,k} \right) -
\lambda\gamma_{k,n,s}^2 p_{k,n,s}\right]
\end{equation}
and $p_{k,n,s}$ is given by~(\ref{eq:solpseudo}). Substituting
back in (\ref{eq:defdualf}), the dual function is
\begin{equation}
  \Theta(\lambda, \boldsymbol{\mu} ) = -\lambda \check{P} +
  \sum_{k=1}^K \mu_k \check{d}_k + \sum_{n=1}^N \min_s f_{n,s} (\lambda, \boldsymbol{\mu} )
  \label{eq:dualf-closed}
\end{equation}
with $f_{n,s}$ given by (\ref{eq:fns-def}). For any value of the
dual variables $(\lambda, \boldsymbol{\mu})$ we can determine the
optimal primal variables $(\boldsymbol{\alpha}, \vect{w})$;
$\boldsymbol{\alpha}$ is obtained by the optimal subcarrier
assignment vector $s(n)$ after performing the minimization over
$s$ in (\ref{eq:dualf-closed}), and the optimal beamforming
vectors $\vect{w}^*_{n,k}$ for the users $k \in s^*(n)$ are given
by
\begin{equation}
\vect{w}^*_{n,k}=\vect{G}^+_{k,n,s^*(n)}[p_{k,n,s^*(n)}^{1/2}, 0, \ldots 0 ]^T.
\end{equation}

The largest part of the computation to evaluate the dual function
is the calculation of $\vect{G}^+_{k,n,s}$ which has to be done
for each subchannel and each possible SDMA set. The number of
evaluations can become quite large but the size of each matrix is
relatively small so that the calculation remains feasible for
medium-size networks. Furthermore, while solving the dual problem
requires multiple subgradient iterations, the calculation of the
pseudo-inverses is \emph{independent} of the value of the
multipliers. This means that the calculation of
$\vect{G}^+_{k,n,s}$ can be done only once in the initialization
step of the subgradient procedure.

\begin{algorithm}
\begin{algorithmic}
\STATE Construct the set $\mathcal{S}$ of all subsets of users of
size $1 \le \kappa \le M$ \FORALL { $n = 1 \ldots N$ } \FORALL {$s
\in \mathcal{S}$} \FORALL {$k \in s$} \STATE Compute  the
pseudo-inverse $G^+_{k,n,s}$ and $\gamma_{k,n,s}$ \ENDFOR
 \ENDFOR
 \ENDFOR
 \STATE Choose an initial value $\lambda^{ 0 } $ and $\boldsymbol{\mu}^{
0  } $ \STATE Subgradient iterations. We set a limit of $I_m$ on
the number of iterations \FORALL {$i = 1 \ldots I_m$} \STATE Solve
the $N$ subproblems (\ref{eq:subproblem}) to compute the dual function
        \STATE Compute the subgradients:
        \STATE $g_{\mu}^{(k)} =  \check{d}_k -  \sum_{n} r_{k,n} $
        \STATE $g_{\lambda} = \sum_{n}  \sum_{k
          \in s^*(n) } \norm{\vect{w}^*_{k,n}  }^2 - \check{P} $
\IF {$\norm{\boldsymbol{g}_{\mu}} \le \epsilon $ and
$\norm{g_{\lambda}} \le \epsilon$}
\STATE {Exit. A dual optimal solution has been found.}
 \ELSE \STATE
Update the multipliers \STATE $\lambda^{i+1}= [\lambda^{i} +
\delta  g_{\lambda} ]^+$ \STATE $\boldsymbol{\mu}^{i+1}=
[\boldsymbol{\mu}^{i} + \delta \boldsymbol{g}_{\mu} ]^+$ \ENDIF
\ENDFOR
\STATE {Exit. A dual optimal solution was NOT found.}
\end{algorithmic}
\caption{Calculation of the Dual Problem} \label{upper-bound-alg}
\end{algorithm}
Note that if the algorithm finds an optimal solution, the
corresponding primal computed from the optimization of the Lagrange
function will be feasible since the subgradients $g_{\mu}^{(k)} $
and $g_{\lambda} $ are small . We use the (somewhat inappropriate)
expression ``dual feasible'' to denote such a solution. If, on the
other hand, the algorithm stops before reaching the optimum, the
primal will generally not be feasible.

Algorithm~\ref{upper-bound-alg} finds the optimal dual variables
$(\lambda^*, \boldsymbol{\mu}^* )$ that solve the dual problem
(\ref{eq:defdualobj}) using the subgradient
method~\cite{bertsekas03} with a fixed step size $\delta$ and
therefore yields an upper bound to problem
(\ref{eq:objzf}--\ref{eq:alphabin}).  In
algorithm~\ref{upper-bound-alg}, the pseudo-inverse matrices
$G^+_{k,n,s}$ can be  computed by any number of well known algorithms.
Here, we have  used the Matlab command \emph{pinv}~\cite{matlab-pinv}. Note that this algorithm can be used to solve
the beamforming problem or equivalently the power allocation
problem for a fixed SDMA set assignment. The only difference is
that~(\ref{eq:subproblem}) becomes trivial since there is only one
possible SDMA set per subcarrier.

It is not possible to evaluate the overall complexity of
algorithm~\ref{upper-bound-alg} because we don't have a bound on the
number of dual iterations. We can nevertheless give a bound on the
complexity of one iteration since it  is
dominated by the pseudo-inverse matrices computation. We have to
compute $NS \sim NK^M$ matrix inversions each one with complexity
$O(M^3)$, giving a total computational complexity $O(N K^M M^3)$.
Computing the subgradient vectors and updating the dual variables
have much lower complexity than the pseudo-inverse matrices
computation.

\subsection{Analysis of the Dual Function}
%
Lets denote $\Theta^*$ the maximum of the dual function
$\Theta(\lambda, \boldsymbol{\mu})$ over $(\lambda,
\boldsymbol{\mu}) \geq 0$. If $U^1$ is the weighted sum-rate
objective function achieved by any feasible point in the primal
problem
 and $U^{*}$ its optimum value, then the following
inequalities hold \cite{boyd04}
\begin{equation}
  \label{eq:ineq-bound}
  U^1 \leq U^* \leq -\Theta^* \leq -\Theta(\lambda, \boldsymbol{\mu})
\end{equation}
The value $-\Theta^*$, or any feasible approximation
$-\Theta(\lambda, \boldsymbol{\mu})$ to it, is thus an \emph{upper
bound} to the optimum value of the primal problem $U^*$. Thus,
$-\Theta^*$ can be used to benchmark any other solution method.

\begin{figure}
\centering
\includegraphics[scale=0.5]{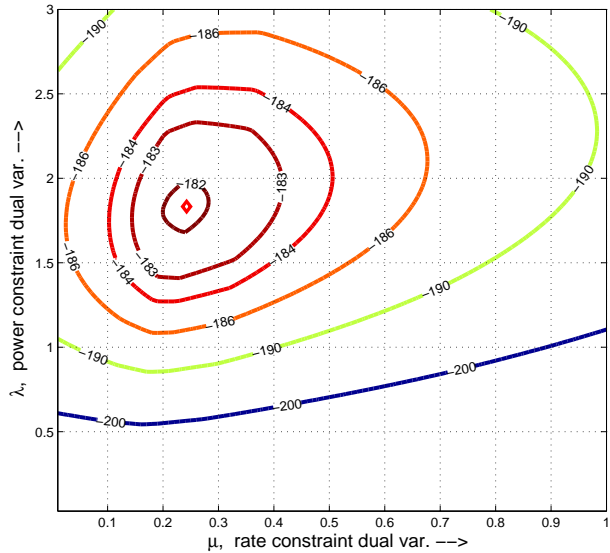}
\caption{Contours of Dual Function, Single RT User. Parameters
$N=8, K=8, M=3, \check{P}=20$ dBm, $\check{d}_1= 50$ bps/Hz}
\label{fig:dualf-contour}
\end{figure}

Figure \ref{fig:dualf-contour} shows a contour plot of the dual
function $\Theta(\lambda, \boldsymbol{\mu})$ for the case of one
RT user. The diamond marker shows the maximum.

\begin{figure}
\centering
\includegraphics[scale=0.5]{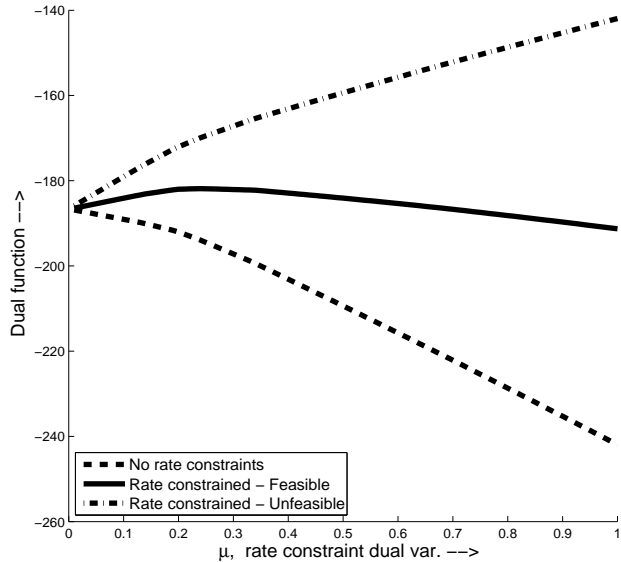}
\caption{Dual Functions for Different Rate Constraints, $\lambda =
1.83$} \label{fig:dualf-mu}
\end{figure}
We can get some insight on the shape of the dual function from
figure \ref{fig:dualf-mu} where we plotted the function with
respect to $\mu$ for a fixed value of $\lambda$. The solid line
curve corresponds to the same dual function as in figure
\ref{fig:dualf-contour}, where the rate constraint is active,
$\check{d}_1= 50$ bps/Hz. We see that the dual function goes
through a maximum at $\mu = 0.24$. We have also shown the case
where we increase the minimum rate constraint so much that the
problem becomes infeasible, e.g., we make $\check{d}_1= 100$
bps/Hz. As expected from duality theory, the dual function has no
maximum since  $ \lim_{\mu \rightarrow \infty} \Theta(\lambda,
\boldsymbol{\mu}) = \infty$ as shown by the dash-dotted curve.
Finally, the dashed curve at the bottom corresponds to
$\check{d}_1= 0$ bps/Hz such that the constraint is inactive and
the solution where the maximum occurs is located at $\mu_1=0$.

\begin{figure}
\centering
\includegraphics[scale=0.5]{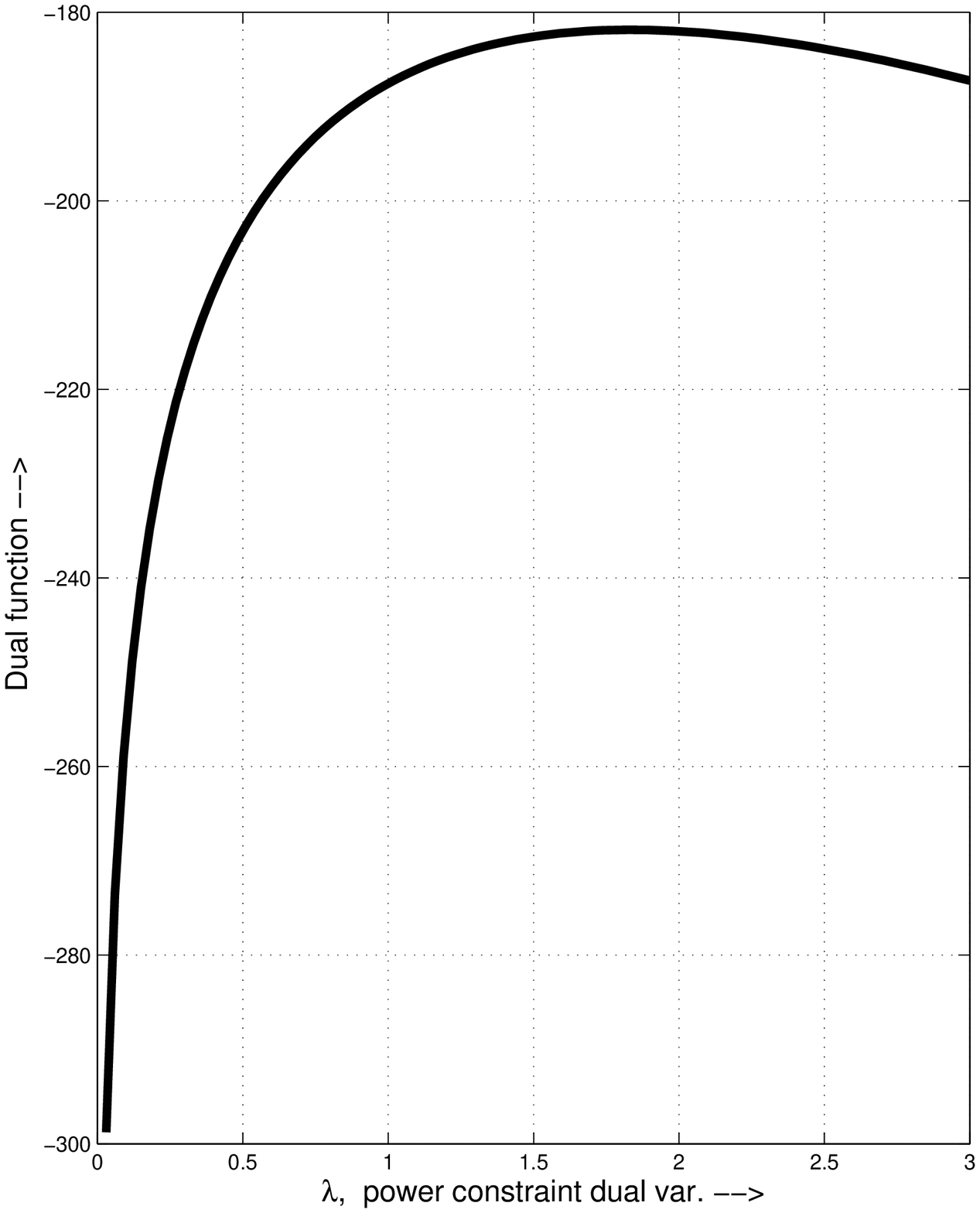}
\caption{Dual Function vs. $\lambda$, $\mu = 0.24$}
\label{fig:dualf-lambda}
\end{figure}
For completeness we show in Figure \ref{fig:dualf-lambda} the dual
function as a function of $\lambda$ when  the rate constraints are
feasible. The dual function increases rapidly and reaches a
maximum at $\lambda = 1.83$.
\subsection{Dual-Based Primal Feasible Method}
\label{sec:find-feas-solut}
The SDMA set selection and beamforming vectors found by
Algorithm~\ref{upper-bound-alg} do not always provide a primal
feasible solution.  The rate or power constraints might be
violated whenever the algorithm stops because the number of
iterations has been reached before the convergence rule is met.
In this subsection,  we propose a simple procedure to obtain a
feasible point to problem (\ref{eq:objzf}--\ref{eq:alphabin}) from
the dual solution found with Algorithm~\ref{upper-bound-alg}. This
point is not optimal but because we start from the dual optimal
solution, we expect that it will be close to the optimal solution. Obviously this
will give us a lower bound to the optimal primal solution.

\begin{algorithm}
\begin{algorithmic}
\STATE { Solve the dual problem~(\ref{eq:defdualobj}) using
algorithm \ref{upper-bound-alg}.  This yields the optimal dual
variables $ \lambda^*, \boldsymbol{\mu}_k^*  $ and a SDMA set
assignment vector ${s}^*(n)$ for each subchannel $n$.}
\STATE{ Set $s^o_0(n) = s^*(n)$} \STATE {Evaluate total power and
user rate constraints (\ref{eq:powerz}--\ref{eq:minratezf})}
\IF{All constraints are met}
   \STATE { Exit. A feasible solution has been found.}
\ENDIF \STATE {Compute power allocation
problem for $s^o_0(n)$ and
evaluate total power and user rate constraints
(\ref{eq:powerz}--\ref{eq:minratezf}) } \IF{All constraints are
met}
   \STATE { Exit. A feasible solution has been found.}
\ENDIF

\STATE {Compute the multipliers $\mu_k$  for users $k$ such that
$r_k < \check{d}_k$}
\FOR {  $j = 1$ to $\overline{J}$  }
       \STATE { $\mu_k = \mu_k + \delta$ }
       \STATE { Find $s^o_j= \arg \min_{s} \{ f_{n,s} \}$

        where $f_{n,s}$ is given by (\ref{eq:theo1_2prove-b}) for the current dual
        variables $\lambda, \boldsymbol{\mu}$ }
        \STATE{ Let $s^o_j(n)$ be the SDMA assignment found}
        \IF { $s^o_j(n) \not= s^o_{j-1}(n)$}
              \STATE { We have found a new SDMA assignment}
              \STATE {Compute power allocation problem for $s^o_j(n)$ and evaluate total power and user rate constraints (\ref{eq:powerz}--\ref{eq:minratezf}) }
                 \IF{All constraints are met}
                    \STATE { Exit. A feasible solution has been found.}
                 \ENDIF

        \ENDIF
 \ENDFOR
 \STATE { Exit. A feasible solution was not found.}

\end{algorithmic}
\caption{Calculating a Feasible Point from the Dual Solution}
\label{algo:dualfeas}
\end{algorithm}
Algorithm~\ref{algo:dualfeas} summarizes this method. The
algorithm begins by solving the dual problem~(\ref{eq:defdualobj})
using Algorithm \ref{upper-bound-alg}. If the solution is not
feasible either directly or by recomputing the  power allocation
for the SDMA set assignment found in the dual problem, the
algorithm performs a search by increasing the dual variables
associated to the users whose QoS constraints are not met until a
new SDMA set assignment is found. It then solves the power
allocation problem for this new SDMA set assignment and checks if the
minimum rate constraints are met.
The search for new SDMA sets continues using this method until a
feasible SDMA set assignment is found or a maximum number of
iteration is reached.

Algorithm~\ref{algo:dualfeas} invokes
algorithm~\ref{upper-bound-alg} which has complexity $O(NK^M
M^3)$. Assuming that the maximum number of iterations $I_M$ is
fixed independently of the problem parameters, the computational
complexity of the feasible point search is $O(NK^M)$ for the
subcarrier reassignment and $O(N(D+K))$ for the power allocation.
These are lower than the computational complexity of algorithm
~\ref{upper-bound-alg}. Therefore, the computational complexity of
algorithm~\ref{algo:dualfeas} is also $O(NK^M M^3)$.

 In contrast
to the enumeration method described in
Section~\ref{subsec:ZF_stra} which performs an enumeration of all
possible SDMA set assignments, the dual-based
Algorithm~\ref{algo:dualfeas} is a method that finds new candidate
SDMA set assignments close to the dual optimal and then uses them
to solve a simple power allocation problem until the rate and
power constraints are met. This makes the search for a
near-optimal feasible point much faster than finding the exact
solution.

\section{Performance Analysis} \label{sec:results}
%

In this section, we present some numerical results on the
performance of the dual-based algorithm and the accuracy of the upper
and lower bounds. To show how they can be used to evaluate heuristic
algorithms, we also compare them with the solution provided by
a weight adjustment method described in Section~\ref{subsec:weight-adjustm-meth}.

%
\subsection{Convergence of the Dual Algorithm}
\label{sec:conv-dual-algor}
\begin{figure}
  \centering
  \includegraphics[scale=0.5]{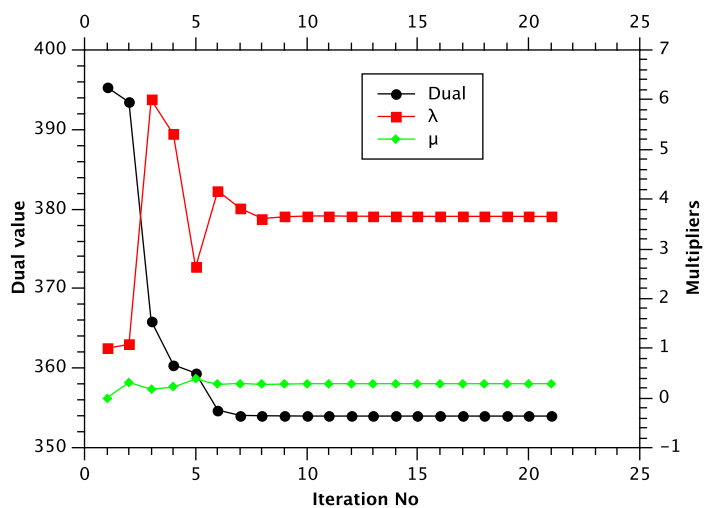}
  \caption{Dual function and multipliers for $M = 3$, $K = 16$, $N = 16$, $\check{P}=20$ dBm, $D=1$ and $\check{d}_1=$80 bps/Hz.}
  \label{fig:dual_min_iter}
\end{figure}

\begin{figure}
  \centering
  \includegraphics[scale=0.5]{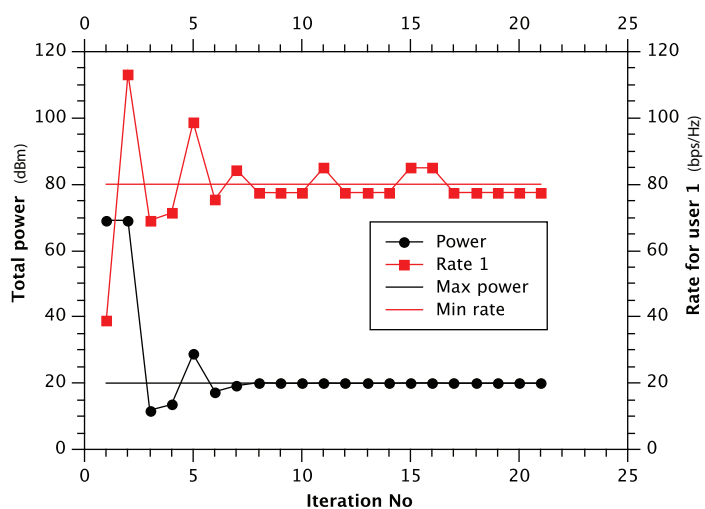}
  \caption{Power and rate constraints for $M = 3$, $K = 16$, $N = 16$, $\check{P}=20$ dBm, $D=1$ and $\check{d}_1=80$ bps/Hz.}
  \label{fig:dual_iter_feas}
\end{figure}

First we present in Figure~\ref{fig:dual_min_iter} the value of
the dual function and Lagrange multipliers as a function of the
number of iterations for a given channel realization. The
corresponding transmit power and the rate received by the RT user
are shown in Figure~\ref{fig:dual_iter_feas}. The parameters used
for the calculation are listed in the figure titles.
We see that the algorithm converges very quickly  to a solution
that is both close to the minimum value and feasible. This is typical
of several other configurations,
except that the number of iteration increases with  the number of RT
users.
\subsection{Weight Adjustment Heuristic}
\label{subsec:weight-adjustm-meth}
As discussed in Section~\ref{sec:state_art}, several RA algorithms
provide support for users with  RT traffic by  increasing the user weights in the
utility function until they receive enough  transmission resources. In this section, we
describe a generic weight adjustment method which will be used to show
that this technique leaves much room for improvement.

%
\begin{algorithm}
\begin{algorithmic}
   \STATE {Solve RA problem (\ref{eq:objzf}--\ref{eq:alphabin}) without minimum rate
constraints constraints (\ref{eq:minratezf}) }
   \STATE { $\vect{c}' \leftarrow \vect{c} $ }
   \STATE { Let $r_k$ be the achieved rate for user $k$ at every
   iteration }
   \STATE { iteration $\leftarrow$ 1 }
   \WHILE { ($r_k < \check{d}_k$ for one or
more users $k \in {\mathcal{D}}$) AND (iteration $\leq  \bar{I}$)
}
       \STATE {Increase user weight using $c_k^{\prime} = c_k^{\prime} +
\epsilon \left( \check{d}_k - r_k  \right) $ for users in need,
where $0< \epsilon \leq 1$  }
           \STATE { Solve RA problem (\ref{eq:objzf}--\ref{eq:alphabin}) without minimum rate
           constraints (\ref{eq:minratezf}) using user weights {$\vect{c}'$}
                \STATE { iteration $\leftarrow$ iteration $+1$ }
} \ENDWHILE
\end{algorithmic}
\caption{Weight Adjustment Algorithm}
 \label{subsubsec:ck-method-alg}
\end{algorithm}
%

We want to find a set of weights
in the utility function~(\ref{eq:objzf}) such that the rate
requirements of the RT users are met when we solve
problem~(\ref{eq:objzf}--\ref{eq:alphabin}) without the rate
constraints~(\ref{eq:minratezf}). Also, the set of weights must
not be very different from one user to the other
in order to maximize the
multi-user diversity gain. Algorithm~\ref{subsubsec:ck-method-alg}
implements a generic method that can do
this. It increases the user weights for RT users until enough
resources are allocated to meet the minimum rate requirements. The
parameter $\epsilon$ controls how much the weights are increased
with respect to the rate bounds.
The rates achieved by
Algorithms~\ref{subsubsec:ck-method-alg} and~\ref{algo:dualfeas} are different since they solve
different problems.
Algorithm~\ref{subsubsec:ck-method-alg} can be seen as solving
problem~(\ref{eq:objzf}--\ref{eq:alphabin}) by using a linear penalty
method for constraints~(\ref{eq:minratezf}) of the form
\begin{displaymath}
  P_k = \min \left\{ 0, r_k - \check{d}_k \right\} .
\end{displaymath}
The modified objective function is then
\begin{align}
  U_P & = \sum_k c_k r_k +  P_k \nonumber \\
  & = \sum_k c_k r_k + \epsilon \sum_{k \vert r_k < \check{d}} (  r_k - \check{d}
  ) . \label{eq:penaltyobj}
\end{align}
At each iteration of the penalty method, whenever rate constraints
are active, the solution of~(\ref{eq:penaltyobj}) cannot be
smaller than that of~(\ref{eq:objzf}--\ref{eq:alphabin}) since it is
a relaxation. Notice that problem (\ref{eq:penaltyobj}) is quite
simple since it has a single constraint for the transmit power but it
has to be solved many times to adjust the weights of the real time
users. In weight adjustment algorithms such as \cite{huang08}, the
user weights are increased  at each time slot using an increasing
function of the packets delay, so the computation task is
distributed over time. However, this distributed approach does not
guarantee that the rate requirements are met in a given time slot
which can lead to delay violations and jitter.

\subsection{Parameter Setup and Methodology}  \label{subsec:results}
We now  present the method and parameter values used to compare
the performance of the different methods to solve problem
(\ref{eq:objzf}). We used a Rayleigh fading model to generate the
user channels such that each component of the channel vectors
$\vect{h}_{k,n}$ are i.i.d. random variables distributed as
$\mathcal{CN}(0,1)$. We also assumed independent fading between
users, antennas and subcarriers. Unless otherwise noted, we used a
configuration with $M=3$ antennas, $K=16$ users and $N=16$
subcarriers. We have only one RT user when we examine the effect of
various parameters and we also look at the impact of increasing the
number of RT users. The minimum rate constraint was set
at $40$ bps/Hz unless otherwise stated. We also fixed the power
constraint to $\check{P}=20$ dBm and used a large-scale
attenuation of 0~dB for all users. The user weights in
(\ref{eq:objzf}) were set to $c_k=1$ for all users. The results
are the average over the feasible cases from 100 independent
channel realizations.

We compared the performance of the different methods for various
scenarios where we increased the resource requirements for the RT
users until the minimum rate requirements can no longer be met for
all RT users. For each scenario and channel realization, the upper
bound was computed  from the dual solution using
Algorithm~\ref{upper-bound-alg} described in
Section~\ref{sec:comp-dual-funct}. For small systems,  we also
found the exact solution using the primal enumeration method given
in Section~\ref{subsec:ZF_stra}. We also computed the lower bound
given by dual-based primal feasible Algorithm~\ref{algo:dualfeas}
and the heuristic solution provided by the weight adjustment
Algorithm~\ref{subsubsec:ck-method-alg} described in
Section~\ref{sec:find-feas-solut}
and~\ref{subsec:weight-adjustm-meth}.  We use the
upper bound given by the dual optimal solution as the reference
point when computing the gap  when the exact solution is not
available.

%
%
\subsection{Single User, Increasing Minimum Rate}
\label{subsec-rmin-results}

\begin{table}
\centering
 \begin{tabular}{ | c |c | c | c |}
 \hline
Method & \multicolumn{3}{|c|}{Minimum rate (bps/Hz)} \\
\cline{2-4}
& $13.33$ &  $16.66$ &  $20$ \\
 \hline\hline
Dual-based upper bound (bps/Hz) & 49.13 & 47.12 & 40.8 \\
 \hline
Primal enum. gap (\%) &  0.57  &  0.55  & 0.10  \\
 \hline
Dual-based feas. gap (\%) &   0.57 &  0.59 &  0.04  \\
 \hline
Weight mod. gap (\%) & 0.68  &  0.71  &  0.15  \\
 \hline
 \end{tabular}
\caption  {Average performance gap against the dual optimal upper
bound for small system configuration.}
 \label{tab:res-comp}
\end{table}

In this first scenario, we have a single RT user and we  increase
its minimum rate $\check{d}_1$. First we consider a small system
with  $K=4$ users and $N=2$ subcarriers where it is possible to
compute an exact solution. We present in table \ref{tab:res-comp}
the average gap in percent between the three methods used to find
feasible solutions against the dual upper bound for a small
system. As the required minimum rate increases from 13.33 to
20~bps/Hz,  the upper bound decreases as more resources need to be
assigned to the RT user until the problem is no longer feasible.
For this small configuration, we see that all methods give
excellent results and the duality gap is very small.

\begin{table}
\centering
\begin{tabular}{ | c |c | c | c | c |}
\hline
 Method & \multicolumn{3}{c|}{Minimum rate (bps/Hz)} \\
\hline
 &  80 &   100 &   120\\
\hline
\multicolumn{4}{|c|}{ Total rate gap against the upper bound (\%)} \\
\hline
Dual-based feas. &   0.24   &  0.23   &  0.21   \\
\hline
Weight mod. &   9.49  &   7.30  &  3.36    \\
\hline
\end{tabular}
\caption {Average total rate gap as a
function minimum rate requirement.} \label{tab:conf2-Rmin}
\end{table}
\begin{figure}
\centering
\includegraphics[scale=0.75]{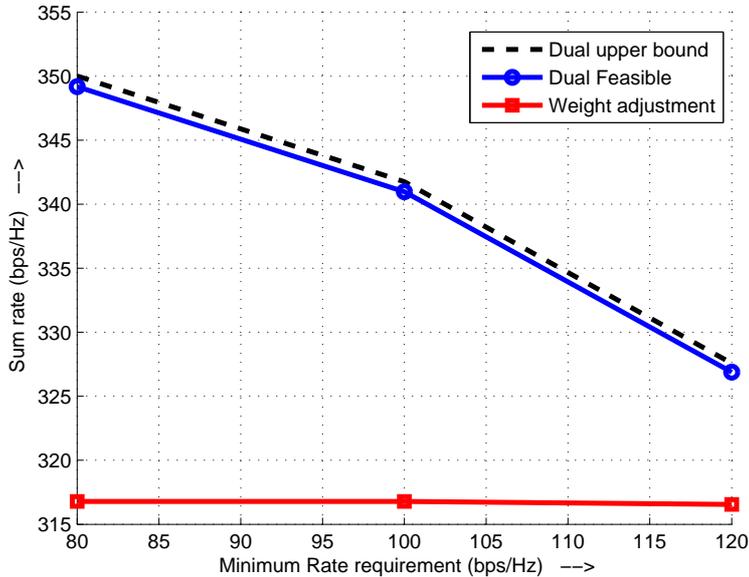}
\caption{Average total rate as a function of the minimum rate
requirements} \label{fig:perf-rmin}
\end{figure}

In the remaining results, we use a larger system with  $K=16$
users and $N=16$ subcarriers. With these values, it is no longer
possible to use the primal enumeration method and we compute the
gap relative to  the dual upper bound. We present in
Table~\ref{tab:conf2-Rmin} the difference in percentage between
the bound  and the solutions of the dual  feasible and the
weight adjustment algorithms. The dual feasible algorithm
gives a lower bound within 0.25\% of the dual upper bound while
the weight adjustment solutions difference can be almost 10\%. As
discussed in Section~\ref{subsec:weight-adjustm-meth}, this is due
to the fact that the weight adjustment algorithm stops as soon as it
finds a feasible solution and does not have the  option of finding
a better assignment. As a result, the
objective does not change much when the minimum rate is increased.
This can be seen from Figure~\ref{fig:perf-rmin} which shows the sum
rate achieved by the dual feasible algorithm and the weight
modification method against the minimum rate requirement.

%
\subsection{Single User, Increasing Attenuation}
\label{sec:single-user-incr}
\begin{figure}
\centering
\includegraphics[scale=0.75]{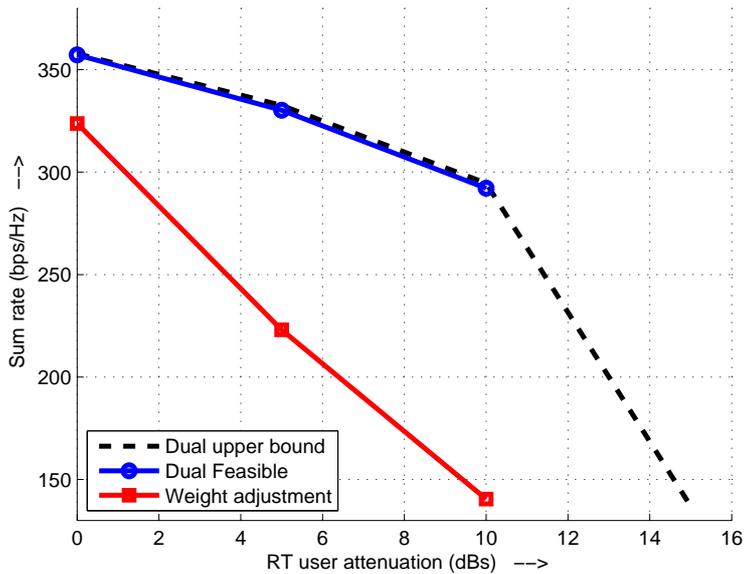}
\caption{Average total rate  as a function of RT user large-scale
channel attenuation.} \label{fig:att-chan}
\end{figure}

As the user moves away from the BS and the channel attenuation
increases, the RA algorithm dedicates more resources to the RT
user until the problem is unfeasible.
Figure~\ref{fig:att-chan} shows the average total rate when the
large-scale channel attenuation of the RT user varies from 0 to 15
dB. The results show that for
all SNR,  the dual lower bound provides a tight solution
with the upper bound while the weight adjustment method shows a
large performance gap.  Table \ref{tab:conf2-SNR} shows  the error
in percentage between the objective and the upper bound. For an
attenuation  of 15 dB, neither method is able to find a feasible
solution; the problem is feasible because the dual upper bound is
around 140, but the algorithms cannot find a solution.  That is,
algorithm~\ref{upper-bound-alg} for the dual upper bound finds a
solution as long as the primal problem is feasible. On the other
hand, algorithm~\ref{algo:dualfeas} cannot find a solution when
the solution point is close to the infeasibility region. In that
case, the algorithm exits declaring that a feasible solution could
not be found. Also, note that the proposed dual-based algorithms
provides a near-optimal to the ZF
problem~(\ref{eq:objzf}--\ref{eq:zfub})  for all SNRs where a
feasible solution can be found by the algorithm.
\begin{table}
\centering
\begin{tabular}{ | c |c | c | c |}
\hline
Method & \multicolumn{3}{c|}{RT user attenuation (dB) } \\
\hline
& 0 &  5 &  10 \\
\hline
\multicolumn{4}{|c|}{ Total rate gap against the upper bound (\%) } \\
\hline
Dual-based feas. &   0.16   &  0.70   &  0.82   \\
\hline
Weight mod. &   9.53  &  32.95  &  52.35    \\
\hline
\end{tabular}\caption {Average total rate gap as a function of RT user large-scale channel attenuation.}
 \label{tab:conf2-SNR}
\end{table}

Figure \ref{fig:att-chan} shows average sum rates. The average is
performed over the channel realizations that provide feasible
points. When the attenuation of the RT user is low, most of the
channel realizations produce feasible points. When it is high, some of the channel realizations do not
produce feasible points and are discarded. For example, in figure~\ref{fig:att-chan} when the attenuation is $15$ dBs, none or very
few  of the channel realizations produce feasible points.
Therefore, figure \ref{fig:att-chan} does not show a sum rate for
that point.
\subsection{Increasing Number of RT Users}
\label{sec:increasing-number-rt}
\begin{figure}
\centering
\includegraphics[scale=0.75]{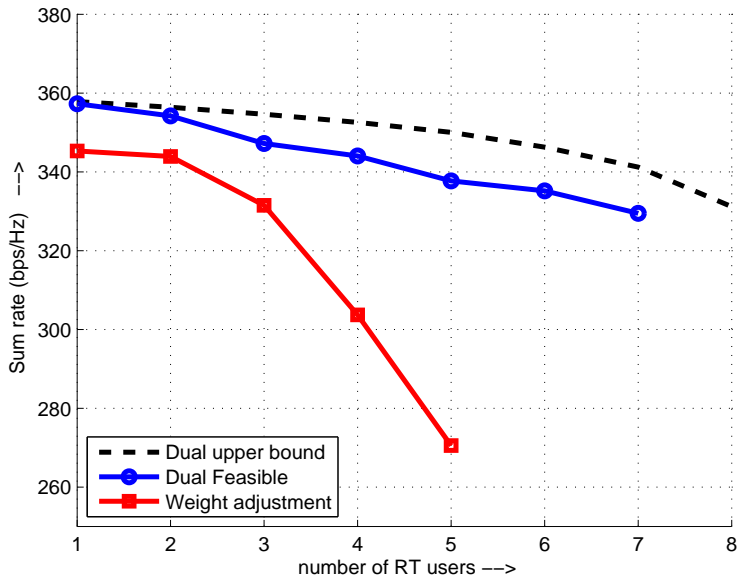}
\caption{Average total rate  as a function of the number of RT
users.} \label{fig:perf-Lusers}
\end{figure}

Finally, figure~\ref{fig:perf-Lusers} shows the upper dual bound,
the lower bound and the solution given by weight adjustment
methods as a function of the number of RT users.
Table~\ref{tab:conf2-Lusers} lists the performance gap against the
dual bound in percentage.   The dual feasible lower bound is again
very close to the upper bound. Meanwhile, we can see that the
performance of the weight adjustment method quickly degrades when
the number of RT users increases.  It cannot find feasible points
with 6 or 7 RT users while the dual algorithm yields solutions for these values  within $3.52 \%$ of the
upper bound.

\begin{table}
\centering
\begin{tabular}{ | c |c | c | c | c |  c | c | c | } \hline
Method & \multicolumn{7}{|c|}{ Number of RT users} \\
\hline
& 1 & 2 & 3 & 4 & 5 & 6 & 7 \\
\hline
& \multicolumn{7}{|c|}{  Total rate gap against the upper bound (\%) } \\
\hline
Dual Feas. &   0.16   & 0.61   &  2.09   &  2.41   &  3.52   &  3.20   &  3.43   \\
\hline
Weight mod. &  3.5    &  3.5   &  6.52    &  13.86    &  22.71    &  - & - \\
\hline
\end{tabular}\caption {Average total rate gap as a function of the number of RT users}
 \label{tab:conf2-Lusers}
\end{table}

For a single RT user, we have seen in tables \ref{tab:conf2-Rmin} and
\ref{tab:conf2-SNR} that the difference between the
upper and lower solution is small. In figure \ref{fig:perf-Lusers}, we
see that this difference increases for three or more
RT users. Still, this growth is not large and
we can consider that a
3.52\% is an acceptable error tolerance. Based on this,  we can claim that it is
possible to find a quasi-optimal solution to problem
(\ref{eq:objzf}--\ref{eq:alphabin}) with the proposed method,
albeit with an off-line algorithm.

Furthermore, the results show that the weight modification method
has a large performance gap which becomes more significant as the
number of  RT users increase. Also, the dual method can find
feasible solutions for cases where the weight adjustment method
cannot. This shows that the weight modification method should be
used carefully for RA in OFDMA-SDMA systems with RT users and that
more efficient heuristics should be developed to approach the
performance of the dual-based feasible solution.
%
\section{Conclusion} \label{sec:conclusions}
In this work, we  proposed a method to compute the beamforming
vectors and the user selection in an OFDMA-SDMA MISO system with
minimum rate requirements for some RT users. We gave a rigorous
mathematical formulation of the Zero-forcing model and then used a
Lagrangian relaxation of the power and rate constraints to solve
the dual problem using a subgradient algorithm.  The Lagrange
decomposition yields sub-problems separated per subcarrier, SDMA
sets and users which substantially reduces the computational
complexity. We obtained a simple expression of the dual function
for the beamforming problem for a given SDMA set based on a
pseudo-inverse condition on the beamforming vectors. The dual
optimum can then be used as a benchmark to compare against any
other solution methods and heuristics. The dual function also
gives us a better understanding of the problem. Its shape is
related to the rate constraint activation and problem feasibility,
and it also justifies the splitting of the subcarrier assignment
and power allocation processes used in several heuristic methods.

We then proposed an algorithm which finds a feasible point  by starting from the
dual-based optimum solution and searching among the dual variables
of the rate constraints. Numerical results indicate that the two
bounds are tight so that the feasible point is near-optimal.
As a point of comparison, we also evaluated the performance of a weight
adjustment method which uses weight adjustments in the objective
function to achieve the required rates. Our results show that
the performance gap of this approach is large and grows
when the SNR of a single RT user increases
or when  the number of RT users increases.

In addition, the weight adjustment
method requires many time slots to adjust the weights and schedule
real time users.  The dual-based method explicitly includes the
minimum rate constraints which allows RT users to be scheduled in
the current slot, which decreases the overall packet delay and
jitter.

The significant gap between the weight adjustment algorithm and
the optimal RA solution, suggests that there is a need to find
better heuristics. The dual approach looks promising to guide the
design of efficient novel heuristics. To implement the RA
algorithm in real time, we also need to design fast methods to
reduce the number of SDMA sets to be searched. The design of these
heuristic algorithms is outside the scope of this paper and is
part of our current efforts. Finally, the upper and lower bounds
also provide a very useful benchmark to compare the performance of
any heuristic method.
\section{Abbreviations}
\begin{description}
\item[BER] Bit Error Rate
\item[BS] Base Station
\item[CPU] Central Processing Unit
\item[LTE] Long Term Evolution  RA Resource Allocation
\item[MIP] Mixed Integer Program
\item[MISO] Multiple Input Single Output
\item[NLP] Non Linear Program
\item[NP] Non-deterministic Polynomial time
\item[nRT] non Real Time
\item[OFDMA] Orthogonal Frequency Division multiplexing Access
\item[QoS] Quality of Service 4G Fourth Generation
\item[RT] Real Time
\item[SDMA] Spatial Division Multiple Access
\item[SISO] Single Input Single Output
\item[SNR] Signal to Noise Ratio
\item[ZF] Zero Forcing
\end{description}
\section{Competing interests} The authors declare that they have no competing
interests.

\begin{acknowledgements}
This research project was  partially supported by NSERC grant
CRDPJ 335934-06.
\end{acknowledgements}

\bibliographystyle{ieeetran}
\bibliography{journal_paper}

\begin{thebibliography}{10}
\providecommand{\url}[1]{#1}
\csname url@samestyle\endcsname
\providecommand{\newblock}{\relax}
\providecommand{\bibinfo}[2]{#2}
\providecommand{\BIBentrySTDinterwordspacing}{\spaceskip=0pt\relax}
\providecommand{\BIBentryALTinterwordstretchfactor}{4}
\providecommand{\BIBentryALTinterwordspacing}{\spaceskip=\fontdimen2\font plus
\BIBentryALTinterwordstretchfactor\fontdimen3\font minus
  \fontdimen4\font\relax}
\providecommand{\BIBforeignlanguage}[2]{{%
\expandafter\ifx\csname l@#1\endcsname\relax
\typeout{** WARNING: IEEEtran.bst: No hyphenation pattern has been}%
\typeout{** loaded for the language `#1'. Using the pattern for}%
\typeout{** the default language instead.}%
\else
\language=\csname l@#1\endcsname
\fi
#2}}
\providecommand{\BIBdecl}{\relax}
\BIBdecl

\bibitem{lte09a}
3GPP, ``{F}urther advancements for {EUTRA}: {P}hysical layer aspects {R}el. 9,
  2010,'' 3{GPP} {TR} 36.814 {V}1.2.1, Tech. Spec.n Group Radio Access Network.

\bibitem{wim09a}
IEEE, ``{D}raft amendment to {IEEE} standard for local and metropolitan area
  networks part 16: {A}ir interface for fixed and mobile broadband wireless
  access systems: {M}ulti-hop relay specification, 2009,'' Standard {IEEE}
  {P}802.16j/{D}9-2009, Institute of Electrical and Electronic Engineers.

\bibitem{letaief06}
K.~Letaief and Y.~Zhang, ``{D}ynamic multiuser resource allocation and
  adaptation for wireless systems,'' \emph{IEEE Wireless Communications
  Magazine}, vol.~13, no.~4, pp. 38--47, Aug. 2006.

\bibitem{bar07}
D.~Bartolom\'e and A.~P\'erez-Neira, ``{P}ractical implementation of bit
  loading schemes for multiantenna multiuser wireless {OFDM} systems,''
  \emph{IEEE Transactions on Communications}, vol.~55, no.~8, pp. 1577--1587,
  Aug. 2007.

\bibitem{maciel07}
T.~F. Maciel and A.~Klein, ``{A} resource allocation strategy for
  {SDMA}/{OFDMA} systems,'' in \emph{Proc. of IST Mobile and Wireless
  Communications Summit}, Jul. 2007, pp. 1--5.

\bibitem{ozbek09}
B.~Ozbek and D.~L. Ruyet, ``{A}daptive resource allocation for {SDMA-OFDMA}
  systems with genetic algorithm,'' in \emph{6th International Symposium on
  Wireless Communication Systems, ISWCS}, Sep. 2009, pp. 483--442.

\bibitem{tsang04}
Y.~Tsang and R.~Cheng, ``{O}ptimal resource allocation in
  {SDMA}/multi-input-single-output/{OFDM} systems under {Q}o{S} and power
  constraints,'' in \emph{Proc. of WCNC}, Mar. 2004, pp. 1595--1600.

\bibitem{chan07}
P.~Chan and R.~Cheng, ``{C}apacity maximization for zero-forcing {MIMO-OFDMA}
  downlink systems with multiuser diversity,'' \emph{IEEE Transactions on
  Wireless Communications}, vol.~6, no.~5, pp. 1880--1889, 2007.

\bibitem{xingmin10}
L.~Xingmin, T.~Hui, S.~Qiaoyun, and L.~Lihua, ``{U}tility based scheduling for
  downlink {OFDMA}/{SDMA} systems with multimedia traffic,'' in \emph{Proc.
  IEEE Wireless Communications and Networking Conference, WCNC}, Mar. 2010, pp.
  130--134.

\bibitem{perea10}
D.~Perea-Vega, J.~Frigon, and A.~Girard, ``{N}ear-optimal and efficient
  heuristic algorithms for resource allocation in {MISO}-{OFDM} systems,'' in
  \emph{IEEE International Conference on Communications ICC}, May 2010, pp.
  1--6.

\bibitem{tsai08}
C.~Tsai, C.~Chang, F.~Ren, and C.~Yen, ``{A}daptive radio resource allocation
  for downlink {OFDMA}/{SDMA} systems with multimedia traffic,'' \emph{IEEE
  Transactions on Wireless Communications}, vol.~7, no.~5, pp. 1734--1743,
  2008.

\bibitem{chung09}
W.~Chung, L.~Wang, and C.~Chang, ``{A} low-complexity beamforming-based
  scheduling for downlink {OFDMA}/{SDMA} systems with multimedia traffic,'' in
  \emph{Proc. of IEEE GLOBECOM}, Nov. 2009, pp. 1--5.

\bibitem{huang08}
W.~Huang, K.~Sun, and T.~Bo, ``{A} new weighted proportional fair scheduling
  algorithm for {SDMA}/{OFDMA} systems,'' in \emph{Proc. 3rd Int. Conf. on
  Communications and Networking in China, ChinaCom}, Aug. 2008, pp. 538--541.

\bibitem{papoutsis10}
S.~K. V.~Papoutsis, I.~Fraimis, ``{U}ser selection and resource allocation
  algorithm with fairness in {MISO-OFDMA},'' \emph{IEEE Communications
  Letters}, vol.~14, no.~5, pp. 411--413, 2010.

\bibitem{vasileios11}
V.~Papoutsis and S.~Kotsopoulos, ``{R}esource {A}llocation {A}lgorithm for
  {MISO-OFDMA} {S}ystems with {Q}o{S} provisioning,'' in \emph{The Seventh
  International Conference on Wireless and Mobile Communications, ICWMC}, 2011,
  pp. 7--11.

\bibitem{tralli11}
V.~Tralli, P.~Henarejos, and A.~Perez-Neira, ``{A} low complexity scheduler for
  multiuser {MIMO}-{OFDMA} systems with heterogeneous traffic,'' in \emph{Proc.
  International Conference on Information Networking, ICOIN}, Jan. 2011, pp.
  251--256.

\bibitem{wang-giannakis08}
X.~Wang and G.~Giannakis, ``{E}rgodic capacity and average rate-guaranteed
  scheduling for wireless multiuser {OFDM} systems,'' in \emph{International
  Symposium on Information Theory, ISIT.}, Jul. 2008, pp. 1691--1695.

\bibitem{yoo06}
T.~Yoo and A.~Goldsmith, ``{O}n the optimality of multiantenna broadcast
  scheduling using zero-forcing beamforming,'' \emph{IEEE Journal on Selected
  Areas in Communications}, vol.~24, no.~3, pp. 528--541, Mar. 2006.

\bibitem{bertsekas03}
D.~Bertsekas, \emph{Convex Analysis and Optimization}.\hskip 1em plus 0.5em
  minus 0.4em\relax Athena Scientific -- Belmont, MA, 2003.

\bibitem{wiesel07}
A.~Wiesel, Y.~Eldar, and S.~Shamai, ``{O}ptimal generalized inverses for zero
  forcing precoding,'' in \emph{Proc. of 41st Annual Conf. on Information
  Sciences and Systems}, 2007.

\bibitem{matlab-pinv}
\BIBentryALTinterwordspacing
Matlab. (2012, Nov.) {M}oore-{P}enrose pseudoinverse of matrix. Matlab
  documentation center. [Online]. Available:
  \url{http://www.mathworks.com/help/matlab/ref/pinv.html}
\BIBentrySTDinterwordspacing

\bibitem{boyd04}
L.~V. S.~Boyd, \emph{Convex Optimization}.\hskip 1em plus 0.5em minus
  0.4em\relax Cambridge University Press, 2004.

\end{thebibliography}

\end{document}